\documentclass{elsart}
\usepackage{amsmath}
\usepackage{epsfig}
\usepackage{hyperref}

\usepackage{graphicx}
\usepackage[usenames]{color}
\usepackage[normalem]{ulem}

\begin{document}
\begin{frontmatter}

\title{Persistent bright solitons in sign-indefinite coupled nonlinear Schr%
\"{o}dinger equations with a time-dependent harmonic trap}

\author{R.Radha$^{\ast,1}$}
\corauth{Corresponding author.}
 \ead{radha\_ramaswamy@yahoo.com, \\
 Telephone: (91)-0435-2403119, Fax: (91)-0435-2403119}
\author{P.S.Vinayagam$^1$}
\author{J.B.Sudharsan$^1$}
\address{$^1$ Centre for Nonlinear Science, PG and Research Dept. of Physics,
Govt. College for Women (Autonomous), Kumbakonam 612001, India}

\author{Boris A. Malomed$^{\dagger,2}$}
\ead{malomed@post.tau.ac.il$^{\dagger }$}
\address{$^2$ Department of Physical Electronics, School of Electrical
Engineering, Faculty of Engineering,Tel Aviv University, Tel Aviv
69978, Israel.}

\begin{abstract}
We introduce a model based on a system of coupled nonlinear Schr\"{o}dinger
(NLS) equations with opposite signs in front of the kinetic and gradient
terms in the two equations. It also includes time-dependent nonlinearity
coefficients and a parabolic expulsive potential. By means of a gauge
transformation, we demonstrate that, with a special choice of the time
dependence of the trap, the system gives rise to persistent solitons. Exact
single- and two-soliton analytical solutions and their stability are
corroborated by numerical simulations. In particular, the exact solutions
exhibit inelastic collisions between solitons.
\end{abstract}
\end{frontmatter}
\newpage

\section{Introduction}

The investigation of multicomponent solitons, which arise due to
the interplay between the second-order dispersion and cubic
nonlinearity, has attracted a great deal of attention, starting
from the classical paper of Manakov \cite{1manakov}, and further
enhanced by works on the copropagation of bimodal waves in
nonlinear optics \cite{2Menyuk}-\cite{7Agrawal}. The dynamics of
multicomponent solitons is described by systems of coupled nonlinear Schr%
\"{o}dinger (NLS) equations [see, e.g., recent works
\cite{8multicomponentNLS}-\cite{10Middelkamp}, and very recent
ones \cite{11SOC}-\cite{17SOC7} dealing with two-component
solitons in spin-orbit-coupled Bose-Einstein condensates (BECs)].
In particular, the concept of energy sharing in the Manakov model
\cite{18manakov-energysharing}, or in the modified version of this
model \cite{19modifiedmanakovmodel} governed by coupled NLS
equations \cite{myprerotation}, was a catalyst for looking for new
integrable models in nonlinear optics \cite{20nonlinearoptics},
BECs \cite{21BEC,22Fei}, metamaterials
\cite{23artificial,24Lazar}, etc. It was found that, in all
available integrable systems of two coupled NLS-type equations,
the ratio between the self-phase-modulation (SPM) and
cross-phase-modulation (XPM) coefficients, which account for the
interaction of the components with themselves or with each other,
are equal, while physically realistic systems depart from this
constraint. Therefore, the quest for new solvable models involving
two coupled NLS equations continues. In this context, Park and
Shin \cite{25ParkShin} have developed new forms of integrable
NLS-type equations going beyond the framework of the conventional
Manakov model, by adding four-wave mixing (FWM) terms to it.

In this paper, we investigate a system of coupled NLS equations
including FWM terms and a time-dependent parabolic potential --
generally, with an anti-trapping sign, i.e., a potential barrier.
An unusual ingredient of the model is that signs in front of the
kinetic and gradient terms are opposite in the two equations.
Hence, it does not directly apply to known physical systems.
Nevertheless, it is interesting as a ``non-standard"
nonlinear-wave model. We employ the gauge-transformation approach
\cite{26llc} to construct bright-soliton solutions of this system.
We conclude that, for a special choice of the trap, bright
solitons persist indefinitely long in the system. We verify the
analytical results by comparing them to the corresponding
numerical simulations, and conclude that the addition of small
perturbations, in the form of sudden variation of the trap's
strength, does not destroy the solitons.

\section{The model}

Waves copropagating in optical media interact through the XPM
nonlinearity \cite{7Agrawal}. Accordingly, the propagation is
governed by the Manakov's model \cite{1manakov} or its
generalization \cite{2Menyuk}-\cite{6Sakaguchi}:

\begin{eqnarray}
iq_{1t}+\frac{1}{2}q_{1xx}+2(g_{11}|q_{1}|^{2}+g_{12}|q_{2}|^{2})q_{1} &=&0,
\notag \\
iq_{2t}+\frac{1}{2}q_{2xx}+2(g_{21}|q_{1}|^{2}+g_{22}|q_{2}|^{2})q_{2} &=&0,
\label{1}
\end{eqnarray}%
where $q_{j}(x,t)$ ($j=1,2$) are envelopes of the field components, $g_{11}$
and $g_{22}$ account for the strengths of the SPM, while $g_{12}$ and $%
g_{21} $ represent the XPM. It is known that eqs. (\ref{1}) are integrable
if either (i) $g_{11}=g_{12}=g_{21}=g_{22}$ or (ii) $%
g_{11}=g_{21}=-g_{12}=-g_{22}$. The former choice corresponds to
the Manakov model proper \cite{1manakov,27manakov-others,{28saha}}
which has been studied in full detail
\cite{18manakov-energysharing,29mana-intensity,30Abdulla}. The
latter choice corresponds to the modified Manakov model
\cite{19modifiedmanakovmodel}, in which the soliton dynamics has
been explored too.

In addition to the XPM, models of the bimodal light propagation in
nonlinear birefringent optical fibers include the FWM terms,
$q_{1}^{2}q_{2}^{\ast }$ and $q_{2}^{2}q_{1}^{\ast }$, which
account for the coherent nonlinear interaction between two linear
polarizations of the electromagnetic waves
\cite{2Menyuk,7Agrawal}. Taking this into regard, we here address
a novel system of coupled NLS-type equations, including the SPM,
XPM, and FWM terms with a time-dependent coefficient, and a
time-dependent anti-trapping parabolic potential. The equations
are written in the notation corresponding to BEC models based on
Gross-Pitaevskii (GP) equations \cite{21BEC,22Fei}:

\begin{eqnarray}
iq_{1t}+\frac{1}{2}q_{1xx}+\gamma (t)(|q_{1}|^{2}-2|q_{2}|^{2})q_{1}-\gamma
(t)q_{2}^{2}q_{1}^{\ast }+\frac{1}{2}\lambda ^{2}(t)x^{2}q_{1} &=&0,  \notag
\\
iq_{2t}+\frac{1}{2}q_{2xx}+\gamma (t)(2|q_{1}|^{2}-|q_{2}|^{2})q_{2}+\gamma
(t)q_{1}^{2}q_{2}^{\ast }+\frac{1}{2}\lambda ^{2}(t)x^{2}q_{2} &=&0,
\label{twogp}
\end{eqnarray}%
where $\gamma (t)$ is the strength of the FWM terms, and $\lambda
^{2}(t)$ is the strength of the anti-trapping (expulsive)
potential. These potentials occur in various physical contexts,
such as the interaction of optical and matter-wave solitons with
barriers \cite{31expulsive}-\cite{33apsvpla2012}, and splitting of
wave packets in interferometers \cite{34Randy}.

Equations (\ref{twogp}) can be derived from the Lagrangian,

\begin{eqnarray}
L &=&\frac{i}{2}\Big(q_{1}^{\ast }\frac{\partial q_{1}}{\partial t}-q_{1}%
\frac{\partial q_{1}^{\ast }}{\partial t}\Big)-\frac{1}{2}\left\vert \frac{%
\partial q_{1}}{\partial x}\right\vert ^{2}+\frac{1}{2}\gamma (t)|q_{1}|^{4}+%
\frac{1}{2}\lambda ^{2}(t)x^{2}|q_{1}|^{2}  \notag \\
&-&2\gamma (t)|q_{1}|^{2}|q_{2}|^{2}-\frac{1}{2}\gamma (t)\Big[%
q_{2}^{2}(q_{1}^{\ast })^{2}+q_{1}^{2}(q_{2}^{\ast })^{2}\Big]-\frac{i}{2}%
\Big[q_{2}^{\ast }\frac{\partial q_{2}}{\partial t}-q_{2}\frac{\partial
q_{2}^{\ast }}{\partial t}\Big]  \notag \\
&+\frac{1}{2}&\left\vert \frac{\partial q_{2}}{\partial x}\right\vert ^{2}+%
\frac{1}{2}\gamma (t)|q_{2}|^{4}-\frac{1}{2}\lambda ^{2}(t)x^{2}|q_{2}|^{2},
\notag
\end{eqnarray}%
with $\ast $ standing for the complex conjugate. An obvious
peculiarity of the Lagrangian is that it is
\emph{sign-indefinite}, as the kinetic and gradient terms, which
contain the $t$- and $x$-derivatives, respectively, feature
opposite signs for components $q_{1}$ and $q_{2}$. For this
reason, this system, with ``opposite directions" of time in the
two subsystems, does not describe currently known physical
settings, although it is somewhat similar to the highly idealized
model of an optical coupler built of normal and
negative-refractive-index cores \cite{35coupler}. Nevertheless,
the system seems quite interesting in its own right, as a
``non-standard" nonlinear-wave model.

Another noteworthy consequence of the opposite ``time
directions" in the two subsystems is nonconservation of the usually defined
total norm,%
\begin{equation}
N=\int_{-\infty }^{+\infty }\left( \left\vert q_{1}(x)\right\vert
^{2}+\left\vert q_{2}(x)\right\vert ^{2}\right) dx.
\end{equation}%
Indeed, a straightforward corollary of eq. (\ref{twogp}) is the following
evolution equation for the norm:%
\begin{equation}
\frac{dN}{dt}=4\gamma (t)\int_{-\infty }^{+\infty }\mathrm{Im}\left\{
(q_{1}^{\ast }(x))^{2}q_{2}^{2}(x)\right\} dx.  \label{dN/dt}
\end{equation}
On the other hand, it is easy to check that the system conserves the \emph{%
difference} between the norms of the two subsystems:%
\begin{equation}
\frac{d}{dt}\int_{-\infty }^{+\infty }\left( \left\vert q_{1}(x)\right\vert
^{2}-\left\vert q_{2}(x)\right\vert ^{2}\right) dx=0,  \label{minus}
\end{equation}%
which is the manifestation of the conservative character of the
system with the ``opposite time directions". In fact,
eq.(\ref{minus}) represents the conservation of energy of the
dynamical system described by eq.(\ref{twogp}). In addition, one
can construct several conserved quantities as in
\cite{36conservative} consolidating the integrability of
eq.(\ref{twogp}). Thus, on the contrary to the \textquotedblleft
normal" systems, where coherent nonlinear coupling leads to
exchange of the norm between the subsystems with the conservation
of the total norm, here the opposite time directions allow the
coherent coupling to generate or absorb the norm.

\section{The Lax pair}
Equations (\ref{twogp}) admits the following Lax-pair
representation:
\begin{eqnarray}
\Phi _{x} &+&U\Phi =0,  \label{Phix} \\
\Phi _{t} &+&V\Phi =0,  \label{Phit}
\end{eqnarray}
where a three-component Jost function is
$\Phi=(\phi_{1},\phi_{2},\phi_{3})^{T}$, and
\begin{eqnarray}
U &=& \left(
\begin{array}{ccc}
i \zeta(t) & P(x,t) K(t) & Q(x,t) K(t)\\
-R1(x,t) K(t)& -i \zeta(t) & 0 \\
-R2(x,t) K(t)& 0 & -i \zeta(t) \\
\end{array}
\right),\label{U}
\end{eqnarray}
\begin{eqnarray}
V&=&\left(
\begin{array}{ccc}
V_{11} & V_{12}& V_{13}\\
V_{21}& V_{22}& V_{23}\\
V_{31}& V_{32} & V_{33}\\
\end{array}
\right),\label{V}
\end{eqnarray}
with
\begin{eqnarray}
V_{11} &=& -i \zeta(t)^2 + i \Omega(t) x \zeta(t) + \frac{i}{2}
\gamma(t) A(t) P(x,t) R1(x,t)K(t)^2 \notag\\
&+& \frac{i}{2} \gamma(t) A(t) Q(x,t)
R1(x,t) K(t)^2 \notag\\
V_{12} &=& (\Omega(t) x -\zeta(t))P(x,t)K(t) + \frac{i}{2} (P(x,t)K(t))_{x}\notag\\
V_{13} &=& (\Omega(t) x -\zeta(t))Q(x,t)K(t) + \frac{i}{2} (Q(x,t)K(t))_{x}\nonumber \\
V_{21} &=& -(\Omega(t) x -\zeta(t))R1(x,t)K(t) + \frac{i}{2}
(R1(x,t)K(t))_{x}\notag\\ V_{22} &=& i \zeta(t)^2 -i \Omega(t) x
\zeta(t)- \frac{i}{2} \gamma(t)
A(t) P(x,t) R1(x,t) K(t)^2\notag\\
V_{23} &=&  -\frac{i}{2} Q(x,t) R1(x,t) K(t)^2,\nonumber \\
V_{31} &=& -(\Omega(t) x -\zeta(t))R2(x,t) K(t) + \frac{i}{2} (R2(x,t)K(t))_{x}\notag\\
V_{32}&=& -\frac{i}{2}P(x,t) R2(x,t) K(t)^2 \nonumber\\
V_{33} &=& i \zeta(t)^2 -i c(t) x \zeta(t)- \frac{i}{2} \gamma(t)
A(t) Q(x,t) R2(x,t) K(t)^2,\nonumber
\end{eqnarray}
with
\begin{eqnarray}
P(x,t)=e^{(i \phi(x,t))}q_1(x,t)\notag\\
Q(x,t)=e^{(i \phi(x,t))}q_2(x,t)\notag\\
R1(x,t)=e^{(-i \phi(x,t))}r_1(x,t)\notag\\
R2(x,t)=e^{(-i \phi(x,t))}r_2(x,t)\notag
\end{eqnarray}
where,
\begin{align}
r_1(x,t)&=-a  q_1(x,t)^{*}- \frac{b  q_1(x,t) q_2(x,t)^{*}}{2 q_2(x,t)} + \frac{d_1  b_1   q_2(x,t)^2 q_1(x,t)^{*}}{2 q_1(x,t)^2 } \notag\\
r_2(x,t)&= \frac{b_1  q_2(x,t)  q_1(x,t)^{*}}{2 q_ 1(x,t)}-c
q_2(x,t)^{*} + \frac{d  b  q_1(x,t)^2 q_2(x,t)^{*}}{2 q_2(x,t)^2 }
\notag\\
A(t)&= \frac{1}{\gamma(t)}; K(t)=\frac{1}{\sqrt{A(t)}};\phi(x,t)=
\Omega (t)x^{2}/2
\end{align}
where $a,b,c,d,b_{1}$ and $d_{1}$ are arbitrary constants. One can
suitably choose these parameters to obtain eq.(\ref{twogp}) as the
compatibility
condition for the Lax pair defined by Eqs. (\ref{Phix})-(\ref{V}), $%
U_{t}-V_{x}+[U,V]=0$, while the spectral parameter $\zeta (t)$ obeys the
following equation:
\begin{equation}
\zeta ^{\prime }(t)=\Omega (t)\zeta (t),  \label{zeta-prime}
\end{equation}%
with
\begin{equation}
\lambda ^{2}(t)=\Omega ^{2}(t)-\Omega ^{\prime }(t),  \label{trap}
\end{equation}%
\begin{equation}
\Omega (t)=-\frac{d}{dt}\ln \gamma (t)  \label{tf}
\end{equation}%
It should be mentioned that Riccati equation.(\ref{trap}) has
already been employed to solve GP-type equations
\cite{37riccatti}-\cite{40VRK}. In fact, the identification of the
Riccati-type equation (\ref{trap}) gives the first
signature of complete integrability of eq. (\ref{twogp}). Equation (\ref%
{trap}), which determines the parabolic-potential strength, $\lambda ^{2}(t)$%
, demonstrates that it is related to the FWM strength, $\gamma (t)$, through
the integrability condition, which can be derived by substituting eq. (\ref%
{tf}) in eq. (\ref{trap}):
\begin{equation}
-\gamma ^{\prime \prime }(t)\gamma (t)+2\left( \gamma ^{\prime }(t)\right)
^{2}-\lambda ^{2}(t)\gamma ^{2}(t)=0.  \label{integ}
\end{equation}%
Thus, the system of coupled GP equations (\ref{twogp}) (or coupled
NLS equations with a time dependent harmonic trap) is completely
integrable for suitable choices of $\lambda (t)$ and $\gamma (t)$,
which are consistent with equation (\ref{integ}). For constant
$\lambda (t)=c_{1}$, eq. (\ref{integ}) yields $\gamma
(t)=e^{c_{1}t}$.

It is worthy to mention that the integrable version of eq. (\ref{twogp}) can
be transformed, by means of substitution%
\begin{equation}
q_{1,2}(x,t)=\frac{1}{\sqrt{\gamma (t)}l(t)}Q _{1,2}(X,T)\exp
(-i\Omega (t)x^{2}/2),
\end{equation}%
with $X\equiv x/l(t)$, $T=T(t)$, $dl/dt=2\Omega l,$ and
$dT/dt=1/l^{2}$, into a system of coupled perturbed NLS equations
with constant coefficients,
\begin{subequations}
\begin{eqnarray}
i\frac{\partial Q _{1}}{\partial T}+\frac{\partial ^{2}Q _{1}}{%
\partial X^{2}}+(|Q _{1}|^{2}-2|Q _{2}|)Q _{1}-Q _{2}^{2}Q
_{1}^{\ast } &=&i\epsilon (t)Q _{1},  \notag \\
i\frac{\partial Q _{2}}{\partial T}+\frac{\partial ^{2}Q _{2}}{%
\partial X^{2}}+(2|Q _{1}|^{2}-|Q _{2}|)Q _{2}-Q _{1}^{2}Q
_{2}^{\ast } &=&i\epsilon (t)Q _{2}.  \label{Kanna}
\end{eqnarray}%
where $\epsilon (t)=(\Omega (t)+\frac{1}{\gamma (t)}\frac{d\gamma (t)}{dt}%
)l^{2}$. Thus, if we choose $\Omega (t)=-\frac{1}{\gamma
(t)}\frac{d\gamma (t)}{dt}$, $\epsilon (t)=0$ and  the above
equation reduces to coherently coupled NLS equation investigated
recently in Ref. \cite {41Kanna} by means of Hirota method.

\section{Persistent solitons and collisional dynamics}

\subsection{Analytical results}

To generate bright vector solitons of eq. (\ref{twogp}), we now consider the
vacuum solution ($q_{1}^{0}=q_{2}^{(0)}=0$), so that the corresponding
eigenvalue problem becomes
\end{subequations}
\begin{eqnarray}
\Phi _{x}^{(0)} &=&U^{(0)}\Phi ^{(0)},  \label{x} \\
\Phi _{t}^{(0)} &=&V^{(0)}\Phi ^{(0)},  \label{t}
\end{eqnarray}%
where
\begin{equation}
U^{(0)}=\left(
\begin{array}{ccc}
-i\zeta (t) & 0 & 0 \\
0 & i\zeta (t) & 0 \\
0 & 0 & i\zeta (t) \\
&  &
\end{array}%
\right) ,  \label{U0}
\end{equation}%
\begin{equation}
V^{(0)}=\left(
\begin{array}{ccc}
-i\zeta (t)^{2}+i\zeta (t)\Omega (t)x & 0 & 0 \\
0 & i\zeta (t)^{2}-\zeta (t)i\Omega (t)x & 0 \\
0 & 0 & i\zeta (t)^{2}-i\zeta (t)\Omega (t)x%
\end{array}%
\right) .  \label{V0}
\end{equation}%
Solving this vacuum linear eigenvalue problem, one gets
\begin{equation}
\Phi ^{(0)}=\left(
\begin{array}{ccc}
e^{-i\zeta (t)x-i\int \zeta (t)^{2}dt} & 0 & 0 \\
0 & e^{i\zeta (t)x+i\int \zeta (t)^{2}dt} & 0 \\
0 & 0 & e^{i\zeta (t)x+i\int \zeta (t)^{2}dt}%
\end{array}%
\right) .
\end{equation}%
We now gauge transform the vacuum eigenfunction $\Phi ^{(0)}$ by a
transformation function $g(x,t)$ to obtain
\begin{eqnarray}
U^{(1)} &=&gU^{(0)}g^{-1}+g_{x}g^{-1},  \label{U1} \\
V^{(1)} &=&gV^{(0)}g^{-1}+g_{t}g^{-1}.  \label{V1}
\end{eqnarray}%
We choose transformation function $g(x,t)$ from the solution of the
associated Riemann problem such that it is meromorphic in the complex $\zeta
$ plane, as
\begin{equation}
g(x,t;\zeta )=\left[ 1+\frac{\zeta _{1}-\zeta _{1}^{\ast }}{\zeta -\zeta _{1}%
}P(x,t)\right] \left(
\begin{array}{ccc}
1 & 0 & 0 \\
0 & -1 & 0 \\
0 & 0 & -1%
\end{array}%
\right) .  \label{g}
\end{equation}%
The inverse of matrix $g$ is given by
\begin{equation}
g^{-1}(x,t;\zeta )=\left(
\begin{array}{ccc}
1 & 0 & 0 \\
0 & -1 & 0 \\
0 & 0 & -1%
\end{array}%
\right) \left[ 1-\frac{\zeta _{1}-\zeta _{1}^{\ast }}{\zeta -\zeta
_{1}^{\ast }}P(x,t)\right] ,  \label{1/g}
\end{equation}%
where $\zeta _{1}$ is an arbitrary complex parameter and $P$ is a
$3\times 3$ projection matrix ($P^{2}=P$) to be determined. The
fact that $U^{(1)}$ and $V^{(1)}$ do not develop singularities
around $\zeta =\zeta _{1}$ and $\zeta =\zeta _{1}^{\ast }$ imposes
the following constraints on $P$:
\begin{eqnarray}
P_{x} &=&(1-P)JU^{(0)}(\zeta _{1}^{\ast })JP-PJU^{(0)}(\zeta
_{1})J(1-P), \\
P_{t} &=&(1-P)JV^{(0)}(\zeta _{1}^{\ast })JP-PJV^{(0)}(\zeta
_{1})J(1-P),
\end{eqnarray}%
where
\begin{equation}
J=\left(
\begin{array}{ccc}
1 & 0 & 0 \\
0 & -1 & 0 \\
0 & 0 & -1%
\end{array}%
\right) .
\end{equation}%
From the above, it is obvious that one can generate projection matrix $P(x,t)
$ using a vacuum eigenfunction, $\Phi ^{(0)}(x,t)$ as $P=J\cdot \tilde{P}%
\cdot J,$ where
\begin{equation}
\tilde{P}=\frac{M^{(1)}}{\mathrm{{Trace}[M^{(1)}]}},
\end{equation}%
\begin{equation}
M^{(1)}=\Phi ^{(0)}(x,t,\zeta _{1}^{\ast })\cdot \hat{m}^{(1)}\cdot \Phi
^{(0)}(x,t,\zeta _{1})^{-1}.  \label{M1}
\end{equation}%
In the above equation, $\hat{m}^{(1)}$ is a $3\times 3$ arbitrary matrix
taking the following form
\begin{equation}
\hat{m}^{(1)}=\left(
\begin{array}{ccc}
e^{2\delta _{1}}\sqrt{2} & \varepsilon _{1}^{(1)}e^{2i\chi _{1}} &
\varepsilon _{2}^{(1)}e^{2i\chi _{1}} \\
\varepsilon _{1}^{\ast (1)}e^{-2i\chi _{1}} & e^{-2\delta _{1}}/\sqrt{2} & 0
\\
\varepsilon _{2}^{\ast (1)}e^{-2i\chi _{1}} & 0 & e^{-2\delta _{1}}/\sqrt{2}%
\end{array}%
\right) ,
\end{equation}%
such that the determinant of $M^{(1)}$ becomes zero under condition $%
|\varepsilon _{1}^{(1)}|^{2}+|\varepsilon _{2}^{(1)}|^{2}=1$. Thus, choosing
$\zeta _{1}=\alpha _{1}+i\beta _{1}$ and using eq. (\ref{M1}), matrix $%
M^{(1)}$ can be explicitly written as
\begin{equation}
M^{(1)}=\left(
\begin{array}{ccc}
e^{-\theta _{1}}\sqrt{2} & e^{-i\xi _{1}}\varepsilon _{1}^{(1)} & e^{-i\xi
_{1}}\varepsilon _{2}^{(1)} \\
e^{i\xi _{1}}\varepsilon _{1}^{\ast (1)} & e^{\theta _{1}}/\sqrt{2} & 0 \\
e^{i\xi _{1}}\varepsilon _{2}^{\ast (1)} & 0 & e^{\theta _{1}}/\sqrt{2}%
\end{array}%
\right) ,
\end{equation}%
where
\begin{eqnarray}
\theta _{1} &=&2x\beta _{1}(t)-4\int (\alpha _{1}(t)\beta _{1}(t))dt+2\delta
_{1}, \\
\xi _{1} &=&2x\alpha _{1}(t)-2\int (\alpha _{1}(t)^{2}-\beta
_{1}(t)^{2})dt-2\chi _{1},
\end{eqnarray}%

with $\left\{ \alpha _{1}(t),\beta _{1}(t)\right\} =\left\{ \alpha
_{10},\beta _{10}\right\} \exp \left( -\int \Omega (t)dt\right) $, while $%
\delta _{1}$ and $\chi _{1}$ are arbitrary parameters.

Now, substituting eqs. (\ref{g}) and (\ref{1/g}) in eq. (\ref{U1}), we
obtain

\begin{equation}
U^{(1)}=\left(
\begin{array}{ccc}
-i\zeta (t) & U^{(0)} & V^{(0)} \\
-U^{(0)\ast } & i\zeta (t) & 0 \\
-V^{(0)\ast } & 0 & i\zeta (t)%
\end{array}%
\right) -2i(\zeta _{1}-\zeta _{1}^{\ast })\left(
\begin{array}{ccc}
0 & \tilde{P}_{12} & \tilde{P}_{13} \\
-\tilde{P}_{12} & 0 & 0 \\
-\tilde{P}_{12} & 0 & 0%
\end{array}%
\right) ,
\end{equation}%
and similarly for $V^{(1)}$. Thus, one can write down the
one-soliton solution as
\begin{eqnarray}
U^{(1)} &=&U^{(0)}-2i(\zeta _{1}-\zeta _{1}^{\ast })\tilde{P}_{12}, \\
V^{(1)} &=&V^{(0)}-2i(\zeta _{1}-\zeta _{1}^{\ast
})\tilde{P}_{13}.
\end{eqnarray}

Thus, the explicit form of one soliton solution can be written as
\begin{eqnarray}
q_{1}^{(1)} &=&2A_{1}\varepsilon _{1}^{(1)}\beta _{0}\mathrm{sech}(\theta
_{1})e^{i(-\xi _{1}+\phi (x,t))},  \label{onesol1} \\
q_{2}^{(1)} &=&2A_{2}\varepsilon _{2}^{(1)}\beta _{0}\mathrm{sech}(\theta
_{1})e^{i(-\xi _{1}+\phi (x,t))},  \label{onesol2}
\end{eqnarray}%
where $\alpha (t),\beta (t)$ are time-dependent scattering lengths and $%
\varepsilon _{1,2}$ are coupling parameters,

with $\phi (x,t)=\Omega (t)x^{2}/2$, $A_{1}=A_{2}=\exp \left[ (1/2)\int
\Omega (t)dt\right] $,
while $\delta _{1}$ and $\chi _{1}$ are arbitrary parameters, and $%
\varepsilon _{1}^{(1)},\varepsilon _{2}^{(1)}$ are coupling constants, which
are subject to constraint $|\varepsilon _{1}^{(1)}|^{2}+|\varepsilon
_{2}^{(1)}|^{2}=1$.

\begin{figure}[tbp]
\begin{center}
\includegraphics[scale=0.55]{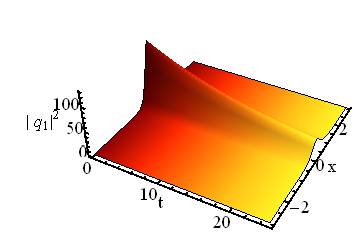}%
\includegraphics[scale=0.55]{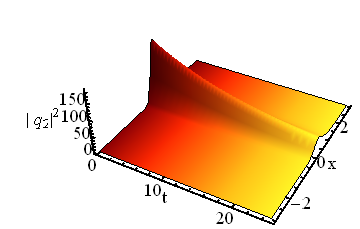}
\end{center}
\caption{(Color online) Decay of the soliton solution in the
time-independent parabolic potential for $\Omega(t) =-0.02$ (or $\gamma(t)=exp [0.02 t]$), $\protect%
\varepsilon _{1}^{(1)}=0.3$, $\protect\alpha _{10}=0.1$, $\protect\beta %
_{10}=0.3$, $\protect\chi _{1}=0.1$, $\protect\delta _{1}=0.2$.}
\label{fig1}
\end{figure}

\begin{figure}[tbp]
\begin{center}
\includegraphics[scale=0.55]{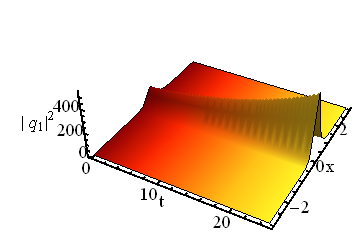}%
\includegraphics[scale=0.55]{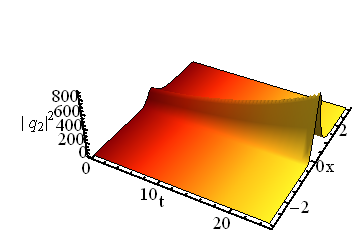}
\end{center}
\caption{(Color online) Growth of the soliton solution for $\Omega(t) =0.02$ (or $\gamma(t)=exp [-0.02 t]$), $%
\protect\varepsilon _{1}^{(1)}=0.3$, $\protect\alpha _{10}=0.5$, $\protect%
\beta _{10}=0.5$, $\protect\chi _{1}=0.5$, $\protect\delta _{1}=0.2$.}
\label{fig2}
\end{figure}

Figures \ref{fig1} and \ref{fig2} show that, in the case of the
time-independent parabolic potential, one observes either decay or
growth of the bright solitons, for a suitable choice of the
potential's strength ($ \Omega (t)=\mathrm{const}$). It should be
also mentioned that the growth and decay of solitons is a
characteristic feature of variable-coefficient NLS-type equations.
For example, the density of the condensates with exponentially
varying scattering length in a parabolic trap grows or decays
\cite{42growthdecay} with time, depending on the sign of
potential, while the underlying dynamical system is completely
integrable and conservative.


\begin{figure}[tbp]
\begin{center}
\includegraphics[scale=0.8]{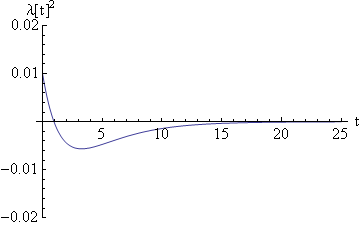}
\end{center}
\caption{(Color online) Evolution of the strength of the parabolic
potential, $\protect\lambda ^{2}(t)$ (which may be both positive and
negative), given by Eq. (\protect\ref{integ}), for $\protect\gamma (t)=\exp %
\left[ \left( 2/3\right) \left( 1-e^{-0.3t}\right) \right] $.}
\label{fig3}
\end{figure}

\begin{figure}[tbp]
\includegraphics[scale=0.55]{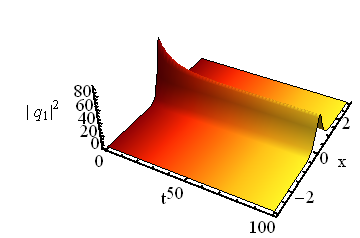}%
\includegraphics[scale=0.55]{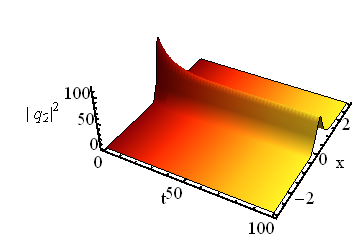}
\caption{(Color online) A persistent soliton for the same $\protect\gamma %
(t) $ as in Fig. \protect\ref{fig3}, and $\protect\varepsilon _{1}^{(1)}=0.3$%
, $\protect\alpha _{10}=0.2$, $\protect\beta _{10}=0.5$, $\protect\chi %
_{1}=0.5$, $\protect\delta _{1}=0.2$.}
\label{fig4}
\end{figure}

To stabilize the solitons, we now introduce a time dependence of the
parabolic potential, selecting $\Omega (t)$ as shown in Fig. \ref{fig3}. For
this case, the density profile of the solution, shown in Fig. \ref{fig4},
indicates that one can sustain the shape of the bright soliton. Accordingly,
we call solutions of the type shown in Fig. \ref{fig4}
``persistent bright solitons".

\subsection{Numerical verification}

\begin{figure}[tbp]
\begin{center}
\includegraphics[scale=0.45]{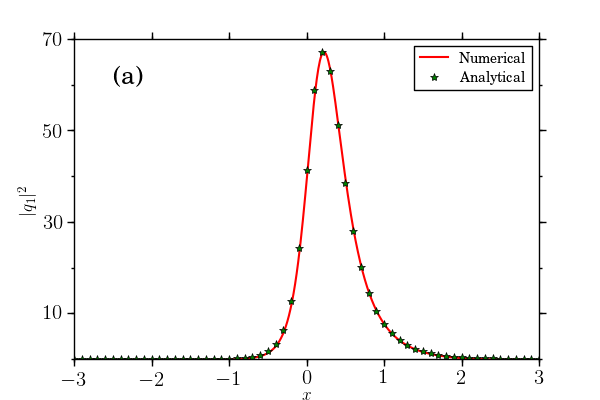}%
\includegraphics[scale=0.45]{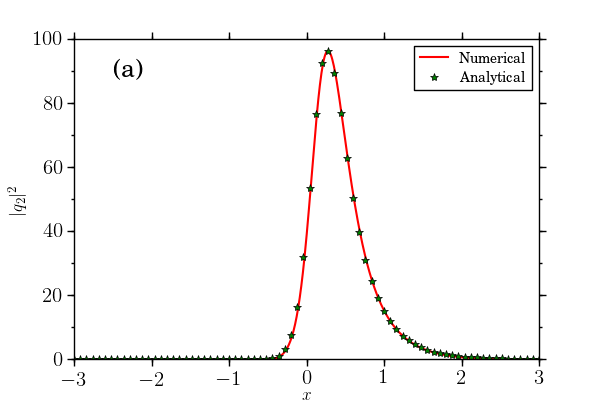}\\[0pt]
\includegraphics[scale=0.45]{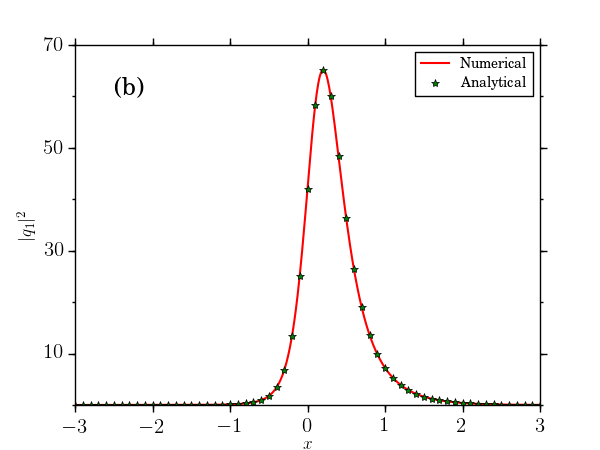}%
\includegraphics[scale=0.45]{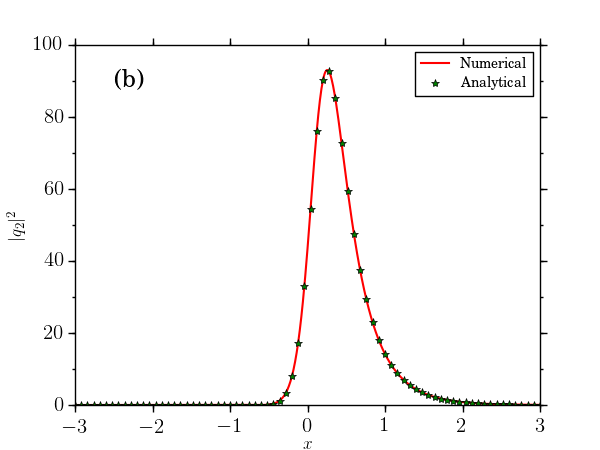}
\end{center}
\caption{(Color online) Comparison of analytical and numerical solutions for
solitons. (a) The upper left panel: the $q_{1}$ component at $t=10$; the
upper right panel: $q_{2}$ at $t=10$. (b) The bottom left panel: $q_{1}$ at $%
t=20$; the bottom right panel: $q_{2}$ at $t=20$. Parameters are the same as
in Fig. (\protect\ref{fig4}). }
\label{fig5}
\end{figure}

It is possible to confirm the analytical results by numerical solutions of
eq. (\ref{twogp}), produced by means of the split-step Crank-Nicolson
method. In Fig. \ref{fig5}, we have plotted the persistent bright solitons
derived analytically as per eqs. (\ref{onesol1}) and (\ref{onesol2}) at $%
t=10 $ and $20$, and the corresponding numerically generated
density profiles. Thus, Fig.(\ref{fig5}) demonstrates exact
matching of the analytical solutions to their numerical
counterparts.

\begin{figure}[tbp]
\begin{center}
\includegraphics[scale=0.65]{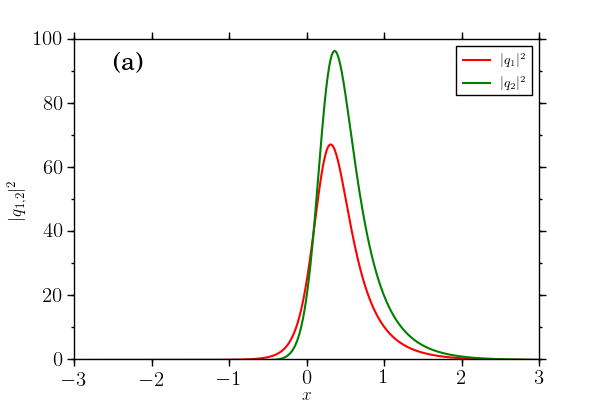}\\[0pt]
\includegraphics[scale=0.65]{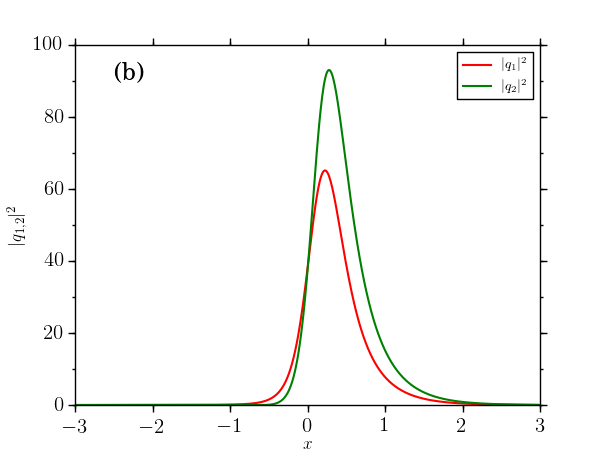}
\end{center}
\caption{(Color online) Density profiles of the fields produced by suddenly
increasing the potential's strength for, $\gamma(t)= exp[(20/3)(1-e^{-0.3t})]$, at (a) $%
t=10$, (b) $t=20$.} \label{fig7}
\end{figure}

\begin{figure}[tbp]
\begin{center}
\includegraphics[scale=0.65]{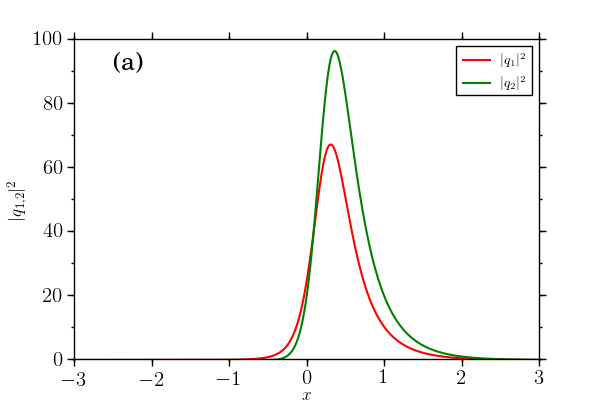}\\[0pt]
\includegraphics[scale=0.65]{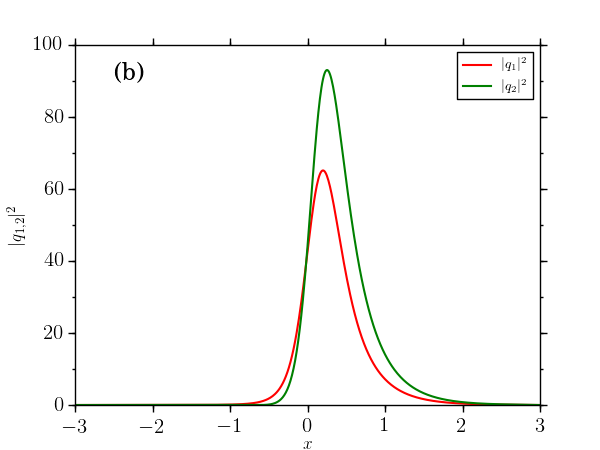}
\end{center}
\caption{(Color online) The density profiles of the fields
obtained while suddenly decreasing the potential's strength, for  $\gamma(t)=exp[(1/15)(1-e^{-0.3t})]$ at (a)$%
t=10$, (b) $t=20$.} \label{fig8}
\end{figure}

Since bright solitons exist due to the special choice of the
strength of the
parabolic potential as a function of time (see eqs. (\ref{trap})-(\ref{integ}%
)), we have also tested the structural stability of the solitons,
by suddenly varying the strength of the potential (either
increasing or decreasing it by $10\%$), as shown in Figs.
(\ref{fig7}) and (\ref{fig8}). From figs.  (\ref{fig7}) and
(\ref{fig8}), we observe that the addition of a small perturbation
does not impact the stability of persistent solitons.

\section{Collisional dynamics of bright vector solitons}

The gauge-transformation approach can be easily extended to generate
multisoliton solutions \cite{26llc}. In particular,, the two-soliton solution $%
q _{1,2}^{(2)}$ for the two modes can be expressed as

\begin{equation}
q_{1}^{(2)}=2IA_{1}/B,~q_{2}^{(2)}=2IA_{2}/B,
\end{equation}%
where
\begin{eqnarray}
A_{1} &=&M_{121}M_{222}\left( \zeta _{2}-\zeta _{1}\right) \left( \zeta
_{1}-\zeta _{1}^{\ast }\right) \left( \zeta _{2}-\zeta _{2}^{\ast }\right)
+M_{122}M_{221}\left( \zeta _{2}-\zeta _{1}^{\ast }\right) \left( \bar{\zeta
_{2}}-\zeta _{1}\right) \left( \zeta _{2}-\zeta _{2}^{\ast }\right)   \notag
\\
&+&M_{111}M_{122}\left( \zeta _{2}-\zeta _{1}^{\ast }\right) \left( \zeta
_{2}^{\ast }-\zeta _{1}^{\ast }\right) \left( \zeta _{2}-\zeta _{2}^{\ast
}\right) +M_{112}M_{121}\left( \zeta _{1}-\zeta _{1}^{\ast }\right) \left(
\zeta _{2}^{\ast }-\zeta _{1}\right) \left( \zeta _{2}^{\ast }-\zeta
_{1}^{\ast }\right) ,  \notag \\
A_{2} &=&M_{112}M_{211}\left( \zeta _{2}-\zeta _{1}\right) \left( \zeta
_{1}-\zeta _{1}^{\ast }\right) \left( \zeta _{2}-\zeta _{2}^{\ast }\right)
+M_{111}M_{212}\left( \zeta _{2}-\zeta _{1}^{\ast }\right) \left( \zeta
_{2}^{\ast }-\zeta _{1}\right) \left( \zeta _{2}-\zeta _{2}^{\ast }\right)
\notag \\
&+&M_{212}M_{221}\left( \zeta _{2}-\zeta _{1}^{\ast }\right) \left( \zeta
_{1}-\zeta _{1}^{\ast }\right) \left( \zeta _{2}-\zeta _{2}^{\ast }\right)
+M_{211}M_{222}\left( \zeta _{1}-\zeta _{1}^{\ast }\right) \left( \zeta
_{2}^{\ast }-\zeta _{1}\right) \left( \zeta _{2}^{\ast }-\zeta _{1}^{\ast
}\right) ,  \notag \\
B &=&\left( M_{122}M_{211}+M_{121}M_{212}\right) \left( \zeta _{1}-\zeta
_{1}^{\ast }\right) \left( \zeta _{2}-\zeta _{2}^{\ast }\right) +\left(
M_{112}M_{221}+M_{111}M_{222}\right) \left( \zeta _{2}-\zeta _{1}^{\ast
}\right)   \notag \\
&&\left( \zeta _{2}^{\ast }-\zeta _{1}\right) +\left(
M_{111}M_{112}+M_{221}M_{222}\right) \left( \zeta _{2}-\zeta _{1}\right)
\left( \zeta _{2}^{\ast }-\zeta _{1}^{\ast }\right) ,  \notag
\end{eqnarray}%

\begin{eqnarray}
M_{11j} &=&e^{-\theta _{j}}\sqrt{2};\quad M_{12j}=e^{-i\xi _{j}}\varepsilon
_{1}^{(j)};\quad M_{13j}=e^{-i\xi _{j}}\varepsilon _{2}^{(j)};  \notag \\
M_{21j} &=&e^{i\xi _{j}}\varepsilon _{1}^{\ast (j)};\quad M_{22j}=e^{\theta
_{j}}/\sqrt{2};\quad M_{23j}=0;  \notag \\
M_{31j} &=&e^{i\xi _{j}}\varepsilon _{2}^{\ast (j)};\quad M_{32j}=0;\quad
M_{33j}=e^{\theta _{j}}/\sqrt{2},  \notag
\end{eqnarray}%
\begin{eqnarray}
\theta _{j} &=&2x\beta _{j}(t)-4\int (\alpha _{j}(t)\beta _{j}(t))dt+2\delta
_{j}, \\
\xi _{j} &=&2x\alpha _{j}(t)-2\int (\alpha _{j}(t)^{2}-\beta
_{j}(t)^{2})dt-2\chi _{1},
\end{eqnarray}%
and $j=1,2$

\begin{figure}[tbp]
\begin{center}
\includegraphics[scale=0.55]{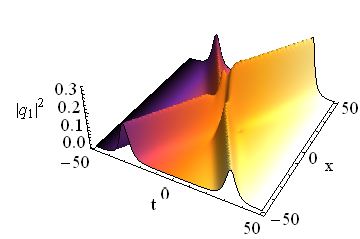}\includegraphics[scale=0.55]{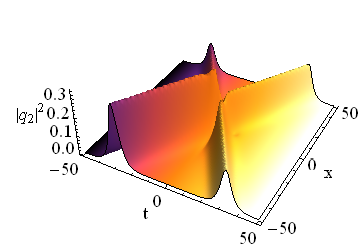}%
\\[0pt]
\includegraphics[scale=0.2]{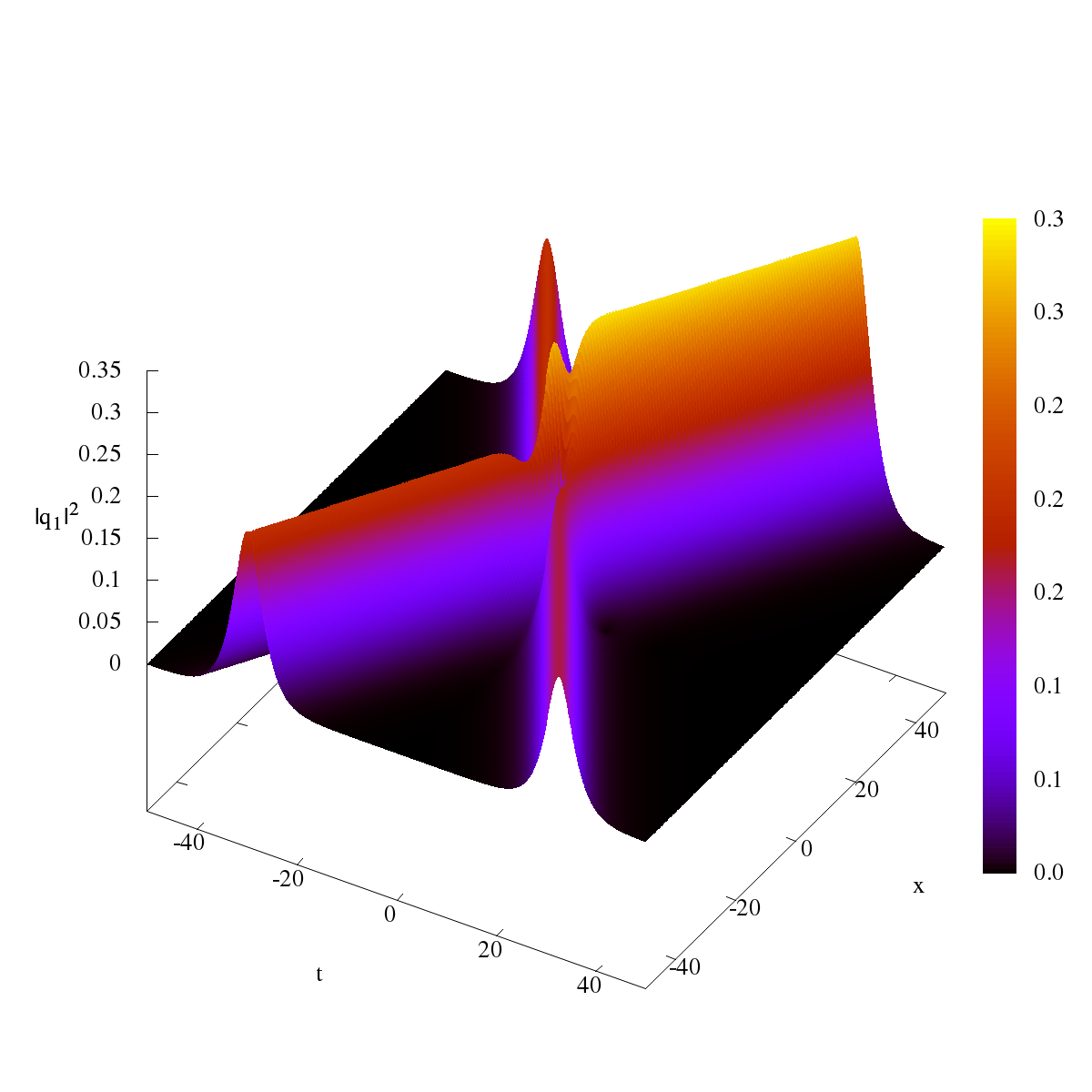}\includegraphics[scale=0.2]{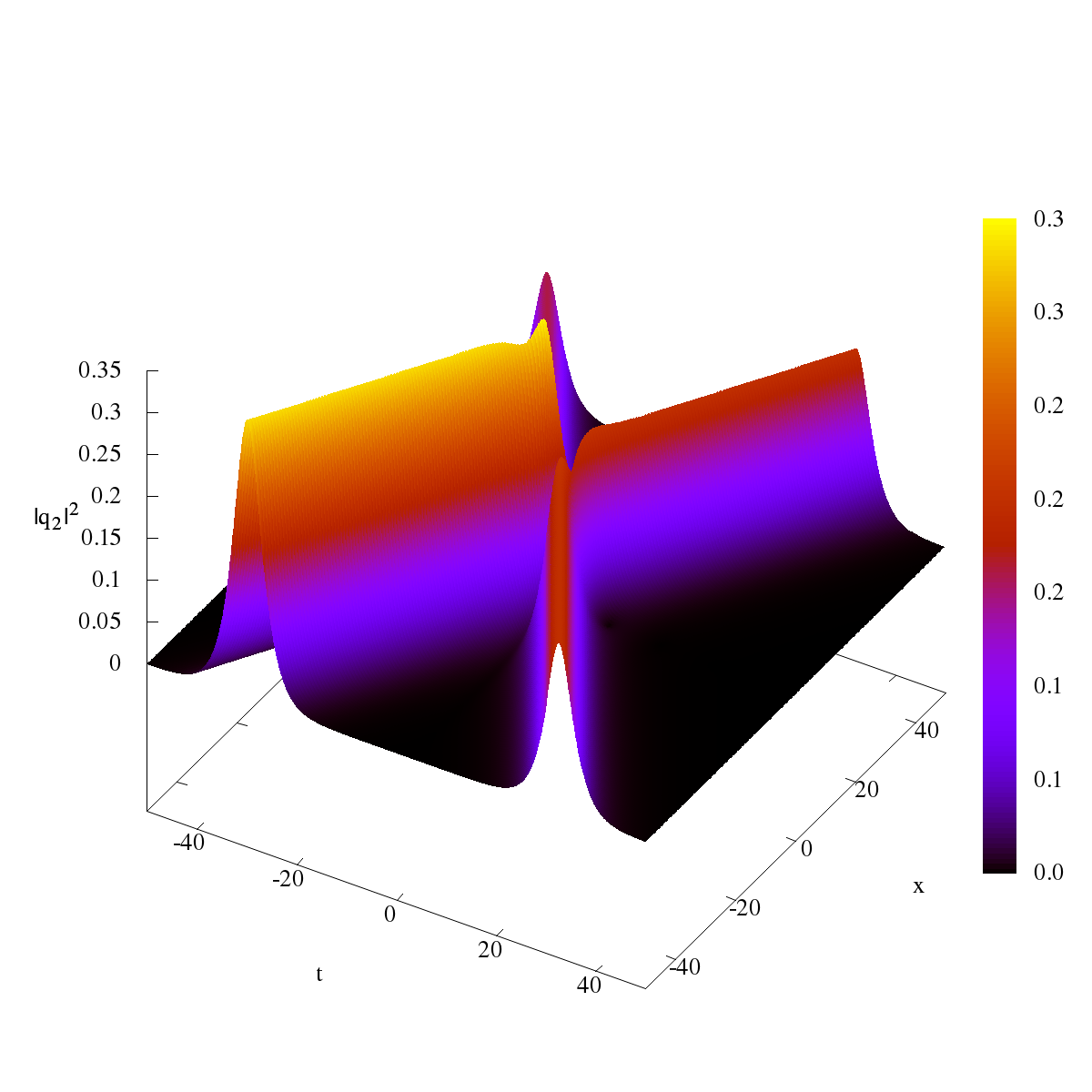}
\end{center}
\caption{(Color online) Inelastic collision of solitons for the choice $%
\protect\gamma (t)=\exp \left[ \left( 2/3\right) \left( 1-e^{-0.3t}\right) %
\right] $., $\protect\alpha _{10}=0.1$, $\protect\alpha _{20}=0.25$, $%
\protect\beta _{10}=0.3$, $\protect\beta _{20}=0.2$, $\protect\delta %
_{1}=0.1 $, $\protect\delta _{2}=0.2$, $\protect\chi _{1}=0.3$, $\protect%
\chi _{2}=0.4 $, $\protect\varepsilon _{1}^{(1)}=0.85i$, $\protect%
\varepsilon _{1}^{(2)}=0.5$such that $|\protect\varepsilon _{1}^{(j)}|^{2}+|%
\protect\varepsilon _{2}^{(j)}|^{2}=1,(j=1,2)$. The top and bottom panels
show the analytical solution and its numerical counterpart.}
\label{twosolfig1}
\end{figure}

In Fig. \ref{twosolfig1}, one can observe inelastic collision of
persistent solitons. The collisional dynamics predicted by the
analytical solution (the top panel in Fig. \ref{twosolfig1}) and
its numerical counterpart (the bottom panel in Fig.
\ref{twosolfig1}) are identical.

\section{Conclusion}

The aim of this work is to investigate the dynamics of solitons in the
integrable system of coupled NLS equations with ``opposite
directions" of time in the two subsystems. The system includes the
time-dependent nonlinearity coefficient, which must be specifically related
to the coefficient in front of the parabolic-potential terms, to secure the
integrability. By means of the gauge transformations, we have demonstrated
that a special choice of the time dependence of the trap may effectively
stabilize bright solitons. We have also observed inelastic collision of
persistent solitons for the same choice of trap frequency which is
subsequently confirmed by numerical simulations.

\section{Acknowledgements}
PSV and JBS thank University Grants Commission (UGC) and
Department of Science and Technology (DST) of India, respectively,
for the financial support. RR wishes to acknowledge the financial
assistance received from DST (Ref.No:SR/S2/HEP-26/2012), UGC
(Ref.No:F.No 40-420/2011(SR), Department of Atomic Energy
-National Board for Higher Mathematics (DAE-NBHM) (Ref.No:
NBHM/R.P.16/2014/Fresh dated 22.10.2014) and Council of Scientific
and Industrial Research (CSIR) (Ref.No: No.03(1323)/14/EMR-II
dated 03.11.2014) dated 4.July.2011) for the financial support in
the form Major Research Projects.

\end{document}